# Data Governance for Platform Ecosystems: Critical Factors and the State of Practice

*Indicate Submission Type: Completed Research Paper*

**Sung Une Lee**[1,2]
Sungune.Lee@data61.csiro.au

**Liming Zhu**[1,2]
Liming.Zhu@data61.csiro.au

**Ross Jeffery**[1,2]
Ross.Jeffery@data61.csiro.au

[1]Architecture and Analytics Platforms Group, Data61, CSIRO, Sydney, Australia
[2]School of Computer Science and Engineering, University of New South Wales, Sydney, Australia

## Abstract

*Recently, "platform ecosystem" has received attention as a key business concept. Sustainable growth of platform ecosystems is enabled by platform users supplying and/or demanding content from each other: e.g. Facebook, YouTube or Twitter. The importance and value of user data in platform ecosystems is accentuated since platform owners use and sell the data for their business. Serious concern is increasing about data misuse or abuse, privacy issues and revenue sharing between the different stakeholders. Traditional data governance focuses on generic goals and a universal approach to manage the data of an enterprise. It entails limited support for the complicated situation and relationship of a platform ecosystem where multiple participating parties contribute, use data and share profits. This article identifies data governance factors for platform ecosystems through literature review. The study then surveys the data governance state of practice of four platform ecosystems: Facebook, YouTube, EBay and Uber. Finally, 19 governance models in industry and academia are compared against our identified data governance factors for platform ecosystems to reveal the gaps and limitations.*

## Introduction

Gawer (2009) defines platforms as the building blocks that act as a foundation upon which an array of firms- sometimes called a business ecosystem or platform ecosystem (our preferred name) can develop complementary products, technologies or services. A platform ecosystem is multi-sided due to different types of firms or user groups transacting with each other (Smedlund & Faghankhani 2015). For example, YouTube has a group which provides videos (side one), and another group watches the videos (side two). The groups facilitate various benefits and grow by providing data by themselves (Evans 2011). This is a so called network effect which is regarded as a key business concept of platform ecosystems. Platform owners (e.g. YouTube) use user data (both user content and non-content like logs or service use history) for their business. It ultimately leads to critical mass and high revenue.

As the ecosystems have been growing rapidly, there are worries of data abuse, privacy violation and proper distribution of profit generated by data. The article titled, "7 Controversial Ways Facebook Has Used Your Data" was reported (Luckerson 2014). It discloses surprising things Facebook can do with user data: e.g. tracking user movements or using user data in ads without consent. According to this article, Facebook paid more than $20 million for lawsuit settlement by disgruntled users. There is also reported danger of data abuse by platform users, partners or employees (Constantin 2011; Tate 2010).

By looking at the example of Facebook, poor implementation or lack of data governance can have significantly destructive effects on success: e.g. losing control of the use of data, lawsuit by disgruntled users and low quality data (Hagiu 2014). Platform owners should overcome these to win markets, and thus to lock the users in the platforms (Parker & Van Alstyne 2014). Meanwhile, platform users need to be guaranteed that their data is safe and transparently managed by the platform owners.

Data governance in the past usually refers to the overall management of the "ilities" (availability, usability, security and privacy) of the data within an enterprise when there is usually clear ownership of data and purpose of use. There is limited research in understanding how data governance should be managed in a platform ecosystem when there are multiple parties contributing, deriving and using data and the ownership, access, usage and profit sharing of collected data and derived data (through data transformation/analysis). We have not been able to find a data governance model for platform ecosystems. Thus, we aim to provide the first study into data governance for platform ecosystems.

The key contributions are threefold. Firstly, this paper assists researchers in understanding data governance in the context of platform ecosystems by conducting an extensive literature review and case studies. The case studies show how data governance is implemented in representative platform ecosystems: Facebook, YouTube, EBay and Uber (the state of practice). Secondly, we highlight data governance factors that should be considered when platform owners implement data governance. In addition, we identify the gaps and limitations of existing governance frameworks for platform ecosystems (state of the art) to discuss future research direction.

In the next section, we present a literature review. We then explain the methodology of this study. Next, we propose our data governance factors for platform ecosystems. Thereafter, we present the state of practice in industry platform ecosystems and the gaps in existing governance models. In the last section, we conclude the paper, and discuss limitations of the study.

## Literature Review

Governance refers to comprehensive control including processes, policies and structures. There are multiple types of governance: e.g. IT/information/data/platform/social network service (SNS) governance. IT governance supports right decision making about IT assets, but data governance focuses on data assets (Khatri & Brown 2010). The term information governance is often used in the same sense as data governance by some authors (Faria et al. 2013), but it focuses on information issues rather than individual data pieces (Dimick 2013). All types of governance are related to each other (Dimick 2013), and share many characteristics (Kamioka et al. 2016). In general, higher level of governance includes lower level of governance as a part. IT governance often addresses information/data governance, and information governance includes data governance (Faria et al. 2013). Thus, lower levels of governance should align with the goals and concepts of higher-level governance. Weill and Ross (2004 and 2005) emphasized that IT governance should align with organizational goals. Khatri and Brown (2010) showed how data governance decision domains align with those of IT governance. In industry, a goal cascading mechanism shows that stakeholder's needs, enterprise goals, IT-related goals and information/data level goals must be aligned (ISACA 2012).

Governance for platform ecosystems (including SNSs) should consider the different business context. Platform ecosystems provide a meeting place, and facilitate interactions between two participating groups (Evans 2011). Platform ecosystems manage the users' data, which is uploaded or generated through these interactions. There is a general consensus in the literature that roles, revenue sharing, trust and control are key governance concepts for platform ecosystems (Schreieck et al. 2016; Hein et al 2016; Tiwana et al. 2010). Roles refer to ownership status (Schreieck et al. 2016). Revenue sharing concepts are focused in prior research as it supports network effects (Schreieck et al. 2016; Parker & Van Alstyne 2014; Tang et al. 2012). Trust between platform owners and users is regarded as a prerequisite factor to success. Control has been addressed in the literature as a vital factor for the use of a successful platform ecosystem (Evans 2011; Ghazawneh & Henfridsson 2010 and 2013; Manner et al. 2012; Tiwana et al. 2010 and 2013).

These governance concepts should be implemented in data governance of platform ecosystems to encourage desirable behavior of all participating groups, and to create value in the use of data (Weber et al. 2009; Kamioka et al. 2016). In addition, the relevant issues and challenges claimed by researchers should be considered. Unclear data ownership (Kaisler et al. 2012 and 2013; Jagadish et al. 2014), the importance of user contribution model (Tang et al. 2012; Parker and Van Alstyne 2014; Chai et al. 2009) and invisible data usage (Kristen 2015) are identified as critical issues. However, in platform ecosystem research, data governance has received little attention (Schreieck et al. 2016).

In some platform governance literature, SNS governance has been identified since SNS is regarded as a representative type of platform ecosystem (Evans 2012). In the literature, focus is also on behavior control of users (Evans 2012), input and output control or privacy violation (Hein et al. 2016) of monitoring and conformance mechanisms. Some of the studies include case studies to present the state of practice of SNSs (e.g. Facebook) in governance structure, control, trust & risk relevant mechanisms. The studies provide good information for understanding real situations, but there is a lack of consideration of how data should be addressed. Web 2.0 has been addressed as an enabling tool for platform ecosystems on the technology level (Prasad et al. 2012) rather than at the (data) governance level.

Research on IT governance provides key decision domains, concepts and principles of governance (Weill & Woodham 2002; Weill & Ross 2004 and 2005) which can be also used in data governance of platform ecosystems. Industry standards or frameworks such as ISO/IEC 38500 or Control Objectives for Information and Related Technologies (COBIT) deliver high-level principles and broad guidance for all size and domain organizations (Chaudhuri 2011; Calder 2008; Sylvester 2011). COBIT, in particular, addresses information/data governance in the framework to support its implementation (Sylvester 2011). Meanwhile, there is less research on information/data governance (Weber et al. 2009; Kamioka et al. 2016). Khatri and Brown (2010) introduced a generic framework to support designing data governance. The framework addresses a broad range of data governance areas, and provides a basic source of practices for data governance for platform ecosystems. Industry information/data governance frameworks provide comprehensive elements to support universal goals of better decision making for companies (Weber et al. 2009). The frameworks provide a goal-driven approach (Begg & Caira 2011) to achieve organizational goals. Thus, those frameworks do not address governance issues crossing organizations or when there are multiple participant types.

In conclusion, there are many academic research and industry governance frameworks. However, there is little research on data governance in the context of platform ecosystems. Furthermore, existing industry frameworks focus on generic information strategy or data management of organizations. In "State of the Art" section, the gaps and limitations of existing industry and research models are shown in detail.

# Methodology

## *Methodology of the Literature Review*

The literature review was performed in three steps following a lightweight systematic literature review: keyword searching, backward and forward searching and literature review based on selection criteria. We used specific query including exchangeable keywords based on Manner et al.'s approach (2012) to look for publications that address platform governance: ((Platform Ecosystem OR multi-sided platform OR two-sided platform) AND (governance OR management)). Through the keyword searching, we found the most recent and relevant three Systematic Literature Review (SLR) papers regarding platform governance (Schreieck et al. 2016; Hein et al 2016; Manner et al. 2012). Backward and forward searching were used to include more literature and to review the latest studies. As a result, we got 231 papers. We filtered them to include suitable papers regarding platform governance, and to exclude those which focus on specific domains or technologies (Table 1). To do so, we first looked at the title to find platform governance studies. We then conducted quick review (looking at the abstract, introduction and conclusion) to confirm whether the topic is really relevant or not. After this, we got 73 papers, and then 16 duplicated papers were removed. Finally, we reviewed the full content of 57 papers and selected papers which explicitly deal with governance concepts, factors or mechanisms in the context of platform ecosystem.

| Reason for inclusion | Reason for exclusion | Review Stage |
|---|---|---|
| Include platform and governance | Not related to platform governance | Title review |
| - | Duplicated literature | Title/quick review |
| Deal with governance concepts, factors or mechanisms in the context of platform ecosystem | Governance concepts/factors/mechanisms are not addressed | Full content review |
| | Focused on specific domains and technologies (e.g. mobile) | |
| | Too high level topic (e.g. overview or strategy) | |
| | Not academic paper | |

**Table 1. The Selection Criteria for Platform Governance Literature**

## *Methodology for the Data Governance Factors*

We focused on both platform governance concepts and current issues/challenges claimed by researchers which should be addressed in data governance for platform ecosystems. We conducted three stages to identify data governance factors and to verify them (Figure 1).

The first stage (step 1~5) was carried out through academic literature review. First of all, we identified four platform governance concepts (Hein et al. 2016; Schreieck et al. 2016) (step 1). We then found six critical issues in prior research which are related to the four concepts, and should be solved for platform ecosystems (Kaisler et al. 2012 and 2013; Jagadish et al. 2014; Tang et al. 2012; Chai et al. 2009) (step 2). In order to find the corresponding data governance factors, we analyzed 14 academic works (step 3). The identified factors were examined if the factors are relevant to the six issues: relevancy (step 4), and if they also are consistent with the four governance concepts: alignment (step 5). Through this process, we identified six data governance factors (F1~6 in Figure 1). The stage 2 (step 3~5) was conducted to verify and refine the factors (and to find missing factors). Five widely used industry governance frameworks were analyzed and examined. We did not find any extra factors from the frameworks, but we could confirm the identified factors in the industry frameworks. As the last stage (step 3~5), we verified and refined the factors by looking at the practices of four platforms: Facebook, Uber, YouTube and EBay. We surveyed good practices in real situations. “Reporting system” is discovered as one type of monitoring mechanisms (for F3, monitoring factor).

Through the three stages, we got six data governance factors, and then added one factor (F7, contribution estimation) to address “Lack of user contribution model” issue and to align “Revenue sharing” concept of platform governance. In total, seven data governance factors were identified.

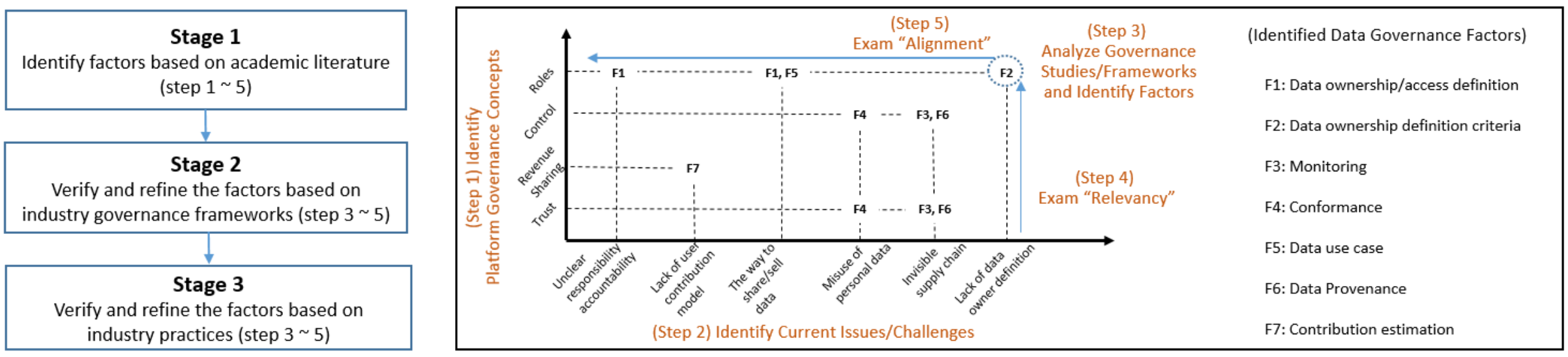

**Figure 1. Research Approach to Identify Data Governance Factors**

## *Methodology of the Case Studies*

The case studies were performed to present the state of practice of platform ecosystems. We categorized platform ecosystems into three types based on Evans’ suggestion (2012): social networks (e.g. Facebook or Twitter), content portals (e.g. YouTube or LinkedIn) and product/service exchanges (e.g. EBay, Uber or Amazon). To decide target platforms, platform’s ranking, size (i.e. active users) and business model were used as the selection criteria. Based on the criteria, Facebook, YouTube and EBay were selected as representative platforms. Uber was additionally chosen as a newcomer for service/product exchanges to compare with the other three large and traditional platform ecosystems.

The studies were conducted by examining their policies: terms and conditions, privacy or data policy (Table 2) because a policy is a central component of governance to represent its concept and to govern the platform ecosystems (ISACA 2012; Schreieck et al. 2016). Furthermore, it is visible and accessible.

| Platform / Category | Facebook (Social Network) | YouTube (Content Portal) | EBay (Product/Service Exchange) | Uber (Product/Service Exchange) |
|---|---|---|---|---|
| Surveyed Policies | - Terms<br>- Data policy<br>- Privacy Basic<br>- Cookies policy | - Terms<br>- Privacy policy<br>- Policy & Safety Center | - Rules & Policies<br>- User Agreement<br>- Cookies<br>- Privacy | - Terms and Conditions<br>- Privacy |

**Table 2. The Surveyed Policies of the Four Platforms**

## Data Governance Factors for Platform Ecosystems

We differentiate between traditional data governance for an enterprise and data governance for a platform ecosystem. We focus on specific governance factors to manage the complicated situation and relationship of a platform ecosystem. Therefore, we do not discuss here all the data governance factors which can appear in a universal data governance model. We identified seven data governance factors based on our approach (Figure 1), and then categorized them into two governance domains (Table 3).

| Domain | Factor | Practice | Reference |
|---|---|---|---|
| Data Ownership /Access | Data ownership and access definition | P1. Define data ownership of all types of data in the platform (user, process and system data)<br>P2. Define access rights based on the ownership and contribution of a data contributor | (Weill & Woodham 2002; Weill & Ross 2004 and 2005; Khatri & Brown 2010; Weber et al. 2009; Otto & Weber 2015; Tiwana et al. 2010 and 2013; Schreieck et al. 2016; Hein et al. 2016) |
| | Definition Criteria | P3. Identify main criteria for defining data ownership<br>P4. Consider relevant regulations (laws, standards and cases)<br>P5. Develop decision models for defining of data owner/access | (Kaisler et al. 2012 and 2013; Weill & Woodham 2002; Weill & Ross 2004 and 2005; Khatri & Brown 2010; Manner et al. 2013) |
| | Contribution Estimation | P6. Consider contributors' effort for value creation<br>P7. Identify dimensions for a measurement model<br>P8. Combine contribution with data ownership/access definition model | (Tang et al. 2012; Parker & Van Alstyne 2014; Chai et al. 2009; Schreieck et al. 2016) |
| Data Ownership /Access; Data Usage | Data Use Case | P9. Define data categories of a platform including various sources (user, process and system data)<br>P10. Define data use cases including individual use case based on the data categories<br>P11. Keep consistency and integrity of the use cases | (Jagadish et al. 2014; Khatri & Brown 2010; Weill & Woodham 2002; Weill & Ross 2004 and 2005) |
| Data Usage | Conformance | P12. Recognize requirements for data due processes<br>P13. Define audit process for conformance of the due processes<br>P14. Consider the result of audit make visible to stakeholders | (Martin 2015; Ghazawneh & Henfridsson 2010 and 2013; Khatri & Brown 2010; Evans 2012; Manner et al 2012 and 2013; Tiwana et al. 2010 and 2013; Schreieck et al. 2016; Hein et al. 2016) |
| | Monitoring | P15. Detect and notify all activities regarding the use of the data in the platform<br>P16. Allow all participating groups to monitor and report the use of the data in platforms<br>P17. Achieve visibility of the data supply chain to stakeholders | (Dempster et al. 2015; Martin 2015; Weill & Woodham 2002; Khatri & Brown 2010; Manner et al 2012 and 2013; Tiwana et al. 2010 and 2014; Schreieck et al. 2016; Hein et al. 2016) |
| | Data Provenance | P18. Enable to trace all the derivation history of the data through metadata management<br>P19. Facilitate data owner authentication through data lifecycle | (Weill & Woodham 2002; Khatri & Brown 2010; Hein et al. 2016) |

**Table 3. The Identified Data Governance Factors for Platform Ecosystems**

(1) Data ownership and access definition: presents who owns and uses the data in platform ecosystems. It has been the focus as a central concept of platform design (Tiwana 2013; Thomas et al. 2014; Schreieck et al. 2016). The definition for all types of data should be clarified to support revenue sharing and to keep control of data flow in platform ecosystems. The definition can be used to protect data against unauthorized use from the platform owners' perspective. Meanwhile, from the users' perspective, it can be regarded as a reward. Responsible, Accountable, Consulted, and Informed (RACI) approach is applicable to define data ownership, but it can be expanded to include "Accessible" for the definition of access rights for platform ecosystems.

(2) Definition criteria: refers to the regulatory environment that could affect the ownership and use of the data in platform ecosystems (Khatri & Brown 2010). Platform users place their data on the Web such as Facebook, YouTube or Twitter. Unclear ownership of the data in the servers entails a serious issue which may have to be resolved in court (Kaisler et al. 2012 and 2013). Therefore, identifying relevant regulations and laws regarding all types of data in platform ecosystems is necessary. Defining an explicit data owner should be implemented according to the identified regulations. A decision model based on legal aspects, and a mechanism to track and notify the compliance of the rules should be applied in data governance for platform ecosystems.

(3) Contribution estimation: is a mechanism to measure user contribution against value creation by providing data. This is necessary to support revenue sharing, to encourage high-quality resources and to repress poor-quality resources of platform ecosystems. For this reason, user contribution measurement refers to a central factor of the business models of platform ecosystems (Tang et al. 2012), and a platform must impose certain regulations on the user participation to reap the benefits of ecosystem growth (Parker & Van Alstyne 2014). However, there is no consideration of user contribution especially in derived and non-content data in research and industry governance models. A contribution measurement model should be adopted for accurately identifying, acknowledging and rewarding the contributors based on the platforms' incentive strategy (Chai et al. 2009).

(4) Data use case: is related to how to use the data in platform ecosystems. Recently, how to share or sell data without losing control has become a central issue to both platform owners and platform users as the value of the data in a platform is increasing. For this, first of all, understanding different types of data is important (Khatri and Brown 2010). Data categories of platform ecosystems should include various sources of data (not only user content): i.e. data from users, systems and business processes (Firmani et al. 2016). In addition, the use cases and the relevant stakeholders of the data (including individual use case for each data category) should be clearly defined and linked. The consistency and integrity of the cases must be considered over the lifecycle of the data to support a visible data supply chain. This factor supports both data ownership and data usage governance domains.

(5) Conformance: is defined as an audit for compliance based on strict processes and rules. Recently, misuse or abuse of personal data has been regarded as the most controversial topic because of a privacy issue. To cope with the issue, minimizing illegal use of identifiable personal data and focusing on data due processes are required (Martin 2015). That is, platform owners should consider conformance checking in their data governance. Apple's review process to assess every application and content against their regulations is a good example (Ghazawneh & Henfridsson 2010). A number of industry governance models and academic research have addressed this as a compulsory aspect.

(6) Monitoring: Data flow should be monitored (Tiwana 2014). Invisible supply chain is a long-standing challenge, but little research has focused on this (Schreieck 2016). A targeted advertising mechanism shows how user data is used through invisible and hidden markets (Craig & John 2015). To overcome this issue, data stewardship practices and a visible supply chain are necessary (Martin 2015). Platform owners should take into account platform users' needs for traceability of who/what has accessed or modified data in a data supply chain (Khatri & Brown 2010). Thus, strict monitoring processes, including reporting by users, should be considered to detect and notify all activities (including internal activities by platform owners and external activities by their partners or the users).

(7) Data provenance: means to trace the derivation history of the data transparently for all participating groups. Metadata is a foundational element to provide transparency and visibility of the use of data (Informatica 2013). Standardizing metadata should be considered for tracking information of the data in platform ecosystems (Khatri & Brown 2010). Data provenance should be retrieved and analyzed through the whole life cycle. In addition, shared data or derived data through transformation or analysis has to be considered for the original data owner (contributor).

## The State of Practice (Case Studies on Platform Ecosystems)

In this section, we present the state of the practice of four platform ecosystems: Facebook, YouTube, EBay and Uber. The purpose of the case studies is to show how the platforms implement data governance. To survey the four platforms, we used seven practices of our data governance factors: P1, P2, P6, P9, P10, P15 and P18 in Tables 3. The practices used here were selected based on visibility in the policies from the platform users' perspective. The practices of "definition criteria" and "conformance" factors were excluded due to invisibility of those practices. Instead, "data provenance" was included to examine whether platform users (data contributors) can trace their data history.

### *The Result Criteria*

We firstly tested if the platforms' documentation shows implementation of the seven practices. If their policies provide information regarding the practices, we conclude the practices satisfy "Existence". We then examined if their data category, ownership, access and data use case definition practices (relevant practices: P1, P2, P9 and P10) address all types of data: "Sufficiency" of the data types. As mentioned in the previous section, we categorized data into three (user content, system and process data) based on the source of the data (Firmani et al. 2016).

Example 1: data ownership practice of Facebook (P1).

- Facebook (their policies) provides information of data ownership definition (i.e. they say who owns the data in the platform in their policies). Thus, "Existence" is satisfied.
- We then looked at the policies in detail to judge if the data ownership definition addresses the three types of data (user, system and process data). Facebook clearly mentions who owns user uploaded data (user content), but does not mention who owns non-content data (i.e. system and process data like logs or service use histories). In this case, "Sufficiency" is not satisfied because Facebook appears to only satisfy one type of data out of three.
- We conclude Facebook "partially implements" the practice (P1) as only "Existence" is satisfied.

Example 2: data ownership practice of EBay (P1).

- We found no information by EBay regarding this practice. Thus, "Existence" is not satisfied.
- Sufficiency is not examined as there is no information to test. Thus, "Sufficiency" is not satisfied.
- We conclude EBay documentation does "not implement" the practice as both criteria are not satisfied.

As above examples, a final result is "Not Implemented", "Partially Implemented" or "Implemented".

## *The Results of the case studies*

- P1. Define data ownership: The existence of "information of data ownership definition" and sufficiency of "collected data types (user content/system data/process data)" are used for this practice. The four platforms' documents do not satisfy the criteria. Facebook, YouTube and Uber define data ownership of user content. However, non-user content (system/process data) is not addressed. EBay documents overlook data ownership.
- P2. Define data access rights: We surveyed the existence of "information of access rights definition" and sufficiency of "collected data types (user content/system data/process data)". Facebook provides detailed information regarding access rights for platform users. However, the other three platforms pay attention to user content only as in P1. Who can access and use non-user content is not clearly mentioned in the documents. Furthermore, it is not clear when platform owners' permission will be terminated. YouTube, EBay and Uber do not provide information regarding data owner's permission withdrawal.
- P6. Consider contributors' effort for value creation: We examined the existence of "information of revenue sharing mechanisms". The documents indicate that users can use their services since they provide their data, and agree with the policies. Information concerning a reward program for a data contributor was not found in the four platforms' policies.
- P9. Define data categories of a platform: The existence of "information of the collected data by the platforms" and sufficiency of "all data types (user content/system data/process data)" are used to survey this practice. All the platforms provide information about "collected data" such as user content, system data (logs or cookies) and process data (service use history). Facebook provides detailed data categories (70 separated categories). YouTube, EBay and Uber present two or three main categories including the detailed sub-categories. This practice is fully documented by all.
- P10. Define general data use cases and individual use cases of each data category: The existence of "purpose of the use of collected data" and sufficiency of "addressed data types (user content/system data/process data)" are surveyed. General purpose of the use of data is well documented by all the platforms: e.g. for better services, communications, safety and security or personalized ads. While YouTube explains the purpose of the use of each data category (e.g. a user profile name and photos for displaying the account information), Facebook, EBay and Uber are not fully satisfied by the criteria. We found the three platforms provide several examples of the purpose of system data such as user cookies or locations. The other types of data were not found.
- P15. Monitor to achieve visibility of the data supply chain: The existence of "information of monitoring mechanisms of the use of data" is used as a key decision criterion to judge if the platforms implement this practice. A limitation in investigating this practice is that we cannot access the whole governance systems of the platforms. The documents show that all platforms do not provide enough capability for platform users to detect the use of their data. They offer report systems to claim abuse of data or services, but there are limitations in seeing transparent data flow.
- P18. Trace all the derivation history of the data in the platforms: The existence of "information of tracking mechanisms and retained data" is important to understand the data provenance systems of the platforms. We investigated how the platforms provide history of the data, including retained data, to the data owner (contributors). However, we observed that this practice is not implemented. We found no information regarding data provenance mechanisms for users in their policies.

We summarized the results of the case studies as follows (Table 4).

| Factor | Surveyed Practices | | | Criteria | Facebook | YouTube | EBay | Uber |
|---|---|---|---|---|---|---|---|---|
| Data Ownership and Access Definition | P1 | **Define data ownership** | | - | △ | △ | ✗ | △ |
| | | Information of Data Ownership Definition | | Existence | o | o | x | o |
| | | Addressed Data Types | User Content | Sufficiency | o | o | x | o |
| | | | System Data | | x | x | x | x |
| | | | Process Data | | x | x | x | x |
| | P2 | **Define data access rights** | | - | ✓ | △ | △ | △ |
| | | Information of Data Access Rights Definition | | Existence | o | o | o | o |
| | | Addressed Data Types | User Content | Sufficiency | o | o | o | o |
| | | | System Data | | o | x | x | x |
| | | | Process Data | | o | x | x | x |
| Contribution Estimation | P6 | **Consider contributors' effort** | | - | ✗ | ✗ | ✗ | ✗ |
| | | Information of Revenue Sharing Mechanisms | | Existence | x | x | x | x |
| Data Use Case | P9 | **Define data categories of a platform** | | - | ✓ | ✓ | ✓ | ✓ |
| | | Information of Collected Data | | Existence | o | o | o | o |
| | | Addressed Data Types | User Content | Sufficiency | o | o | o | o |
| | | | System Data | | o | o | o | o |
| | | | Process Data | | o | o | o | o |
| | P10 | **Define data use cases** | | - | △ | ✓ | △ | △ |
| | | Information of Data Use Case | | Existence | o | o | o | o |
| | | Addressed Data Types | User Content | Sufficiency | x | o | x | x |
| | | | System Data | | o | o | o | o |
| | | | Process Data | | x | o | x | x |
| Monitoring | P15 | **Monitor the data supply chain** | | - | ✗ | ✗ | ✗ | ✗ |
| | | Information of data usage monitoring mechanisms | | Existence | x | x | x | x |
| Data Provenance | P18 | **Trace the derivation history of the data** | | - | ✗ | ✗ | ✗ | ✗ |
| | | Information of Tracking Mechanisms | | Sufficiency | x | x | x | x |

✓ Implemented △ Partially Implemented ✗ Not Implemented o Satisfied x Not Satisfied

**Table 4. The Results of the Case Studies**

### *Discussion*

We identified some significant issues in data governance of the four platforms. Firstly, we found a lack of consideration of various sources of data. The platforms focus on user content, but it is not clearly defined who owns or uses non-user content (e.g. logs, keywords or purchase history). As stated previously, various types of data should be considered in every governance decision domain. The owner of data should be clearly identified to reduce uncertainty and argument between participating groups. Secondly, data flow in the supply chain is not visible to platform users. The platforms' policies are imprecise, and thus how, when, and who uses the data are not clear. This issue is claimed in previous research as one of the critical challenges and ethical issues (Kaisler et al. 2013; Martin 2015). This issue should be resolved to keep trust between platform owners and the users, and to succeed in the business (Schreieck et al. 2016; Hein et al. 2016). To overcome this, monitoring and data provenance systems should be strengthened. Lastly, absence of consideration for data contributors is identified as one of the critical issues. To encourage network effects, user participation should be rewarded as platform users always expect immediate rewards or future benefits (Tang et al. 2012).

## State of the Art (Existing Governance Models)

We analyzed currently used industry and research governance models to reveal gaps and limitations based on our identified data governance factors (Table 3). To do so, we categorized governance areas into three parts: IT governance, information/data governance, and governance for platform ecosystems (including SNS governance). We reviewed widely recognized or used frameworks of IT governance and data governance. For instance, COBIT was selected because it is the best known and most widely used framework (Weber et al. 2009). The Weill IT governance, the Khatri and Brown, and the Weber et al.'s model were also included as their studies have been widely referenced by other researchers in the area. For the platform governance studies category, we included studies which address platform governance approaches, concepts or factors based on our criteria mentioned in the methodology section. Eight platform governance studies were finally chosen.

## *The Result Criteria*

As in the previous section, we used “Sufficiency” to decide if the models have concern of the suggested data governance factors. For this, we used all practices of each factor (Table 3). A final result was determined “Not Covered”, “Partially Covered” or “Covered” based on the coverage of the practices.

## *Results of the Analysis*

The result was tabulated to show the summary of the analysis (Table 5).

| Governance Category/Framework (rows) \ Platform Data Governance Factors (columns) | | Ownership and Access Definition | Definition Criteria | Contribution Estimation | Data Use Case | Conformance | Monitoring | Data Provenance |
|---|---|---|---|---|---|---|---|---|
| IT Governance | ISO/IEC 38500 | ◐ | ✗ | ✗ | ✗ | ◐ | ◐ | ✗ |
| | COBIT 5.0 | ◐ | ◐ | ✗ | ◐ | ● | ◐ | ✗ |
| | Weill and Woodham (2002) | ◐ | ◐ | ✗ | ◐ | ✗ | ◐ | ✗ |
| | Weill and Ross (2004 and 2005) | ◐ | ◐ | ✗ | ◐ | ✗ | ✗ | ✗ |
| Information /Data Governance | DGI Framework | ● | ✗ | ✗ | ✗ | ✗ | ◐ | ✗ |
| | Informatica Framework | ● | ✗ | ✗ | ✗ | ● | ● | ● |
| | IBM Information Governance Component Model | ◐ | ◐ | ✗ | ◐ | ◐ | ◐ | ◐ |
| | Khatri and Brown (2010) | ◐ | ◐ | ✗ | ● | ◐ | ◐ | ◐ |
| | Weber et al. (2009) and Otto and Weber (2015) | ◐ | ✗ | ✗ | ✗ | ✗ | ✗ | ✗ |
| Governance for Platform Ecosystems (*Literature which includes SNS governance) | Evans (2012)* | ✗ | ✗ | ✗ | ✗ | ◐ | ◐ | ✗ |
| | Ghazawneh and Henfridsson (2010 and 2013) | ✗ | ✗ | ✗ | ✗ | ◐ | ✗ | ✗ |
| | Hagiu (2014)* | ✗ | ✗ | ✗ | ✗ | ✗ | ✗ | ✗ |
| | Manner et al. (2012) | ✗ | ✗ | ✗ | ✗ | ◐ | ◐ | ✗ |
| | Manner et al. (2013) | ✗ | ◐ | ✗ | ✗ | ◐ | ◐ | ✗ |
| | Tiwana et al. (2010) | ◐ | ✗ | ✗ | ✗ | ◐ | ◐ | ✗ |
| | Tiwana (2013) | ◐ | ✗ | ✗ | ✗ | ◐ | ◐ | ✗ |

● Covered ◐ Partially Covered ✗ Not Covered

**Table 5. The Results of an Analysis of Existing Governance Models**

There is no industry IT governance framework or studies that address “contribution estimation” and “data provenance”. In contrast, responsibility and accountability (data ownership) are emphasized by the surveyed models. ISO/IEC 38500 and COBIT stresses clarity of role, responsibility and accountability through RACI definition (Chaudhuri 2011; ISACA 2012). Weill’s studies with Woodham (2002) and Ross (2004 and 2005) focus on the importance of IT (data) emphasizing the role of IT to meet the business goal. “Definition criteria” for data ownership is also considered as a part of IT architecture concern in their studies. However, the five industry frameworks and research focus on traditional concerns of responsibility and accountability within an enterprise. Thus, there is a critical lack of access rights concern from platform users’ perspective. “Conformance and Monitoring” factors are partially covered by the frameworks. Yet, there is no concern of data supply chain visibility.

In information/data governance models, as in the IT governance category, “contribution estimation” is not considered. “Data provenance” is regarded as one area of major concern by the Informatica framework (Informatica 2013). The framework describes that data lineage is necessary to support metadata of critical data and to provide transparency and visibility of the flow of data. The IBM information governance model also address this factor. The model, however, focuses on organizational data flow through transformations and processing of data (Ballard et al. 2014), rather than platform ecosystem context. The Khatri and Brown model also includes data provenance concern (Khatri & Brown 2010). The model aims to provide governance design concepts. Thus, it doesn’t draw attention to the detailed and accurate study of the factor. “Data ownership/access definition” is covered by all the surveyed information/data governance models. Industry models, in particular, focus on this factor. The DGI framework fully covers this in the core components (Thomas 2006). Access rights control is described as a role of Data Governance Office (DGO). The Informatica framework also mentions data ownership and access with concern for transparent and clear policies (Informatica 2013). However, the instruction level is very high. Weber et al. (2009) and Otto and Weber (2015)'s studies partially cover data ownership/access concern by focusing on assignment of roles to tasks to define clear responsibilities. Their typical data governance approach for an in-house control mechanism is recognized as the limitation. That is, external stakeholders (i.e. platform users) are not considered in the studies. "Data access rights" of stakeholders is similarly not involved. "Data use case" factor is addressed by two frameworks. The IBM model addresses this concern by emphasizing the role of data steward for all information/data sources (Ballard et al. 2014). Yet, there is a lack of relationship between data use case and the relevant stakeholders. In contrast, the Khatri and Brown model includes all the practices of the factor. However, there is a limitation to use in practice because the abstraction level is high. The importance of "conformance and monitoring" is emphasized in the Informatica framework by mentioning compliance with the organization's requirements (Informatica 2013). However, other frameworks do not cover or partially cover the factors. The DGI framework presents "monitoring" as one of the role of data governance office, but how to achieve and maintain visibility of a data supply chain is missing.

Platform governance studies mainly concentrate on "conformance and monitoring" factors. Evans (2012) has an interest in how platform ecosystems can govern bad behavior by platform users. Meanwhile, Ghazawneh and Henfridsson (2010 and 2013) addressed regulation-based securing to control a platform ecosystem through boundary resources such as APIs, documents or data. They had a focus on a conformation mechanism such as due processes of administrative legislation. Manner et al. (2012 and 2013) also introduced control mechanisms which include monitoring and a conformance concept. Tiwana et al. (2010 and 2013) addressed output and process control as a "conformance" concept to ensure desirable outcomes and verify compliance. "Monitoring" of user behaviors in accordance with prescribed methods and criteria is also emphasized. Data ownership and access are also discussed with incentives issues (Tiwana et al. 2010; Tiwana 2013). The discussion, however, is not sufficient to clearly explain who owns and how to use the data. "Definition criteria" for data ownership is introduced by Manner et al. (2013). They spotlighted the fact that access to the data in platform ecosystems by participants is increasing. Their concern, however, is less focused on how to share the results of monitoring, conformance and data ownership definition.

### *Discussion*

By looking at the results, all existing frameworks and previous studies have disregarded data providers' contribution. Moreover, access rights issues based on user's contribution are also mostly neglected. The importance of data provenance is also not addressed in most frameworks. A limited number of frameworks or studies have an interest in data flow management. A majority of frameworks focus on how to control platforms through conformance and monitoring mechanisms. The studies overlook the importance of visibility of a data supply chain.

To sum up, there are some critical points which should be strengthened in existing governance models: (1) how to achieve visibility and traceability of a platform ecosystem, (2) how to combine traditional RACI method and access rights of data contributors, and (3) how to measure contribution effort of platform users.

## Conclusion and Limitations

In this paper, we proposed seven data governance factors for platform ecosystems, and presented the state of practice through case studies and state of the art to show current gaps and limitations.

To identify the factors, we reviewed academic literature, and then verified and refined them by using industry frameworks and practices. The identified factors align with platform governance concepts (roles, revenue sharing, control and trust) to differentiate from traditional data governance. In addition, we focused on what data governance domains are currently an issue. Data ownership and access rights of platform users are not clearly defined. In particular, non-user content is overlooked. Moreover, there is lack of mechanisms to support visibility of data flow. Data contributor's effort is also not addressed in any existing governance frameworks and industry practices. The case studies were conducted by surveying the policies of four platform ecosystems to show how the platforms document data governance practices. We then showed the gaps and limitations of 19 industry and research governance models by comparing these with our data governance factors.

Future research is planned to study efficient and effective ways of implementing the suggested factors. To do so, we need to overcome several limitations that remain in this study. First of all, our data governance factors need to be analyzed based on different views (e.g. platform owners vs platform users) in practice. Secondly, our case studies have possible validity issues. The case studies were carried out and assessed by one author, and limited data sources were used. We need to consider validation of our observations, and multiple data sources for analysis. In addition, we need to choose various types of platform ecosystems such as C2C, B2C or B2B platforms to improve diversity of sampling, and to strengthen external validity and generalization. Lastly, our literature search might not include all relevant studies. We need to expand the search boundary to include more literature by using a complete SLR approach.